\documentclass[aps,prl,twocolumn,showpacs,showkeys,footinbib,amsmath,amssymb,superscriptaddress]{revtex4-2}

\usepackage{amsmath}
\usepackage{amssymb}
\usepackage{graphicx}
\usepackage{bm}
\usepackage{color}
\usepackage{relsize}
\usepackage{braket}
\usepackage{bbold}
\usepackage{txfonts}
\usepackage{epstopdf}
\PassOptionsToPackage{colorlinks=true,
	linkcolor=blue,
	citecolor=blue,
	urlcolor=blue}{hyperref}
\usepackage{hyperref}

\begin{document}

\title{Altermagnetoelectric Spin Field Effect Transistor}

\author{Ziye Zhu}
\affiliation{Eastern Institute for Advanced Study, Eastern Institute of Technology, Ningbo, Zhejiang 315200, China}

\author{Xianzhang Chen}
\affiliation{Eastern Institute for Advanced Study, Eastern Institute of Technology, Ningbo, Zhejiang 315200, China}

\author{Xunkai Duan}
\affiliation{Eastern Institute for Advanced Study, Eastern Institute of Technology, Ningbo, Zhejiang 315200, China}

\author{Zhou Cui}
\affiliation{Eastern Institute for Advanced Study, Eastern Institute of Technology, Ningbo, Zhejiang 315200, China}

\author{Jiayong Zhang}
\affiliation{Eastern Institute for Advanced Study, Eastern Institute of Technology, Ningbo, Zhejiang 315200, China}
\affiliation{School of Physical Science and Technology, Suzhou University of Science and Technology, Suzhou, 215009, China}

\author{Igor \v{Z}uti\'c}
\affiliation{Department of Physics, University at Buffalo, State University of New York, Buffalo, New York 14260, USA}

\author{Tong Zhou}
\email{tzhou@eitech.edu.cn}
\affiliation{Eastern Institute for Advanced Study, Eastern Institute of Technology, Ningbo, Zhejiang 315200, China}

\date{\today}

\begin{abstract}
Spin field-effect transistors (SFETs) are promising candidates for low-power spin-based electronics, yet existing realizations that rely on spin–orbit coupling are constrained by limited material choices and short spin-coherence lengths. Here we propose a different operating principle based on multiferroic altermagnets, in which spin splitting is tuned by an electric field through symmetry control rather than conventional spin–orbit physics. Using an effective model combined with quantum transport simulations, we show that the conductance is determined by the degree of matching between the electrically controlled spin texture of the channel and the fixed spin polarization of ferromagnetic contacts, enabling clear ON and OFF states. Remarkably, we also address a long-standing challenge in multiferroic device design: spintronic channels require metallic carriers, whereas ferroelectricity is usually suppressed in metals. We resolve this conflict by imprinting multiferroic altermagnetism into highly conductive materials via the proximity effect. First-principles calculations for graphene on multiferroic vanadium sulfide halides confirm that graphene acquires a ferroelectrically switchable spin splitting while retaining its metallic character. These results establish a practical route to SFET implementation and identify multiferroic altermagnets as a versatile platform for next-generation spintronic devices.
\end{abstract}

\maketitle
Spintronics exploits the electron spin to achieve energy-efficient information processing, with phenomena such as spin polarization and spin–orbit coupling (SOC) forming the basis of key device concepts~\cite{Zutic2004RMP, Chappert2007NM, Hirohata2020JMMM}. Among them, the spin field-eﬀect transistor (SFET), proposed by Datta and Das~\cite{Datta1990APL}, had a profound influence on concepts in spintronics and on the exploration of devices beyond magnetoresistance, despite the challenges for its experimental implementation. Traditional SFET designs rely on SOC-induced spin splitting in narrow-gap semiconductors, enabling the modulation of spin precession through gate voltages~\cite{Datta1990APL}. However, the scarcity of suitable materials and issues such as spin decoherence have significantly hindered their robustness and practical realization~\cite{Cao2023Nature}. These challenges have fueled efforts to achieve spin control through electric fields, a strategy that promises enhanced robustness and lower energy consumption compared to SOC-based modulations~\cite{Matsukura2015NN, Gong2025NN}. 

The recent discovery of altermagnets (AMs), which combine key advantages of both ferromagnets and antiferromagnets, has opened new avenues for spintronics~\cite{Smejkal2022PRX1, Smejkal2022PRX2, Bai2024AFM, Song2025NRM, bhowal2025arxiv, Cheong2025npj,Wu2007PRB, Yuan2020PRB, Libor2020SciAdv, JunWeiLiu2021NC}. A wide range of phenomena and device concepts have been proposed in these systems~\cite{Chen2025PRL, Chen2025Nature, Hulunhui2025PRL, Chenwei2025PRB, Liu2025Arxiv, Bandyopadhyay2025PRB, GuoSDPRB2025, Shao2023PRL, Liu2025PRB, Sun2025PRB, Yang2025PRB, Chi2025PRApplied, Guo2023npj, Zhu2023PRB, Li2023PRB, Ghorashi2024PRL, Sun2025Arxiv, Cao2025PRL, KTLaw2025Arxiv, RuanJiawei2025Arxiv, Libo2025Arxiv, Lizhe2025Arxiv, Yanzhongbo2025Arxiv, Zhu2025Arxiv, Duan2024PRL, Gu2024PRL, Smejkal2024Arxiv, Zhu2025NanoLetters, Zhu2025SCPMA, SunWei2025Arxiv, Cao2024Arxiv, Bhowal2025PRL, Guo2025Arxiv, UrruPRB2025, Sun2025AM, Peng2025Arxiv, Huang2025PRL, Ding2025Arxiv, Gougaoyang2025Arxiv}, including altermagnetic tunnel junctions~\cite{Shao2023PRL, Liu2025PRB, Sun2025PRB, Yang2025PRB, Chi2025PRApplied}, topological states~\cite{Guo2023npj, Zhu2023PRB, Li2023PRB, Ghorashi2024PRL}, excitonic effects~\cite{Sun2025Arxiv, Cao2025PRL}, and proximity effect~\cite{Zhu2025Arxiv}. Among them, multiferroic altermagnets~\cite{Duan2024PRL, Gu2024PRL, Smejkal2024Arxiv, Zhu2025NanoLetters, Zhu2025SCPMA, SunWei2025Arxiv, Cao2024Arxiv, Bhowal2025PRL, Guo2025Arxiv, UrruPRB2025, Sun2025AM, Peng2025Arxiv, Huang2025PRL, Ding2025Arxiv, Gougaoyang2025Arxiv}, where (anti)ferroelectric order coexists with altermagnetism, offer a particularly promising route for fast and reversible electrical control of magnetism and spin polarization, making them natural candidates for SFET applications. A central feature of these materials is that the magnetoelectric coupling does not arise from a direct interaction between electric and magnetic order parameters. Instead, ferroelectricity modifies the crystal symmetry, and the symmetry-sensitive altermagnetic order adjusts accordingly. This symmetry-driven multiferroicity can greatly enhance the effective magnetoelectric response and thereby enable highly efficient electric control of magnetic properties~\cite{Duan2024PRL, Gu2024PRL, Smejkal2024Arxiv, Zhu2025NanoLetters, Zhu2025SCPMA, SunWei2025Arxiv}.

Despite these advances, a practical device architecture that exploits multiferroic altermagnets has not yet been demonstrated. For device applications such as SFETs, metallic channels with mobile carriers are essential, but multiferroic altermagnets are typically insulating in order to sustain ferroelectricity. This intrinsic mismatch between the need for metallic transport and the usually insulating nature of multiferroic altermagnets poses a serious obstacle to device integration~\cite{Lines2001book}.

To address these challenges, we propose a new spintronic device: the altermagnetoelectric spin field-effect transistor (AMSFET). Its operation relies on symmetry-mediated locking between ferroelectric polarization and altermagnetic spin texture in multiferroic altermagnets, which enables fully electrical ON/OFF control of spin transport. Using a tight-binding model combined with quantum-transport simulations, we demonstrate the universality and robustness of this operating principle. To reconcile the need for metallic transport with the typically insulating nature of ferroelectric multiferroics, we transfer the multiferroic altermagnetic order into highly conductive channels via the proximity effect. First-principles calculations for graphene on multiferroic vanadium sulfide halides show that graphene develops a ferroelectrically switchable spin splitting while preserving its metallic character. These results establish the viability of AMSFETs and identify multiferroic altermagnets as a new class of channel materials for spintronics, advancing SFET concepts and helping to resolve the long-standing tension between ferroelectricity and electronic conduction.

\begin{figure}[t]
\includegraphics
[width=0.48\textwidth]{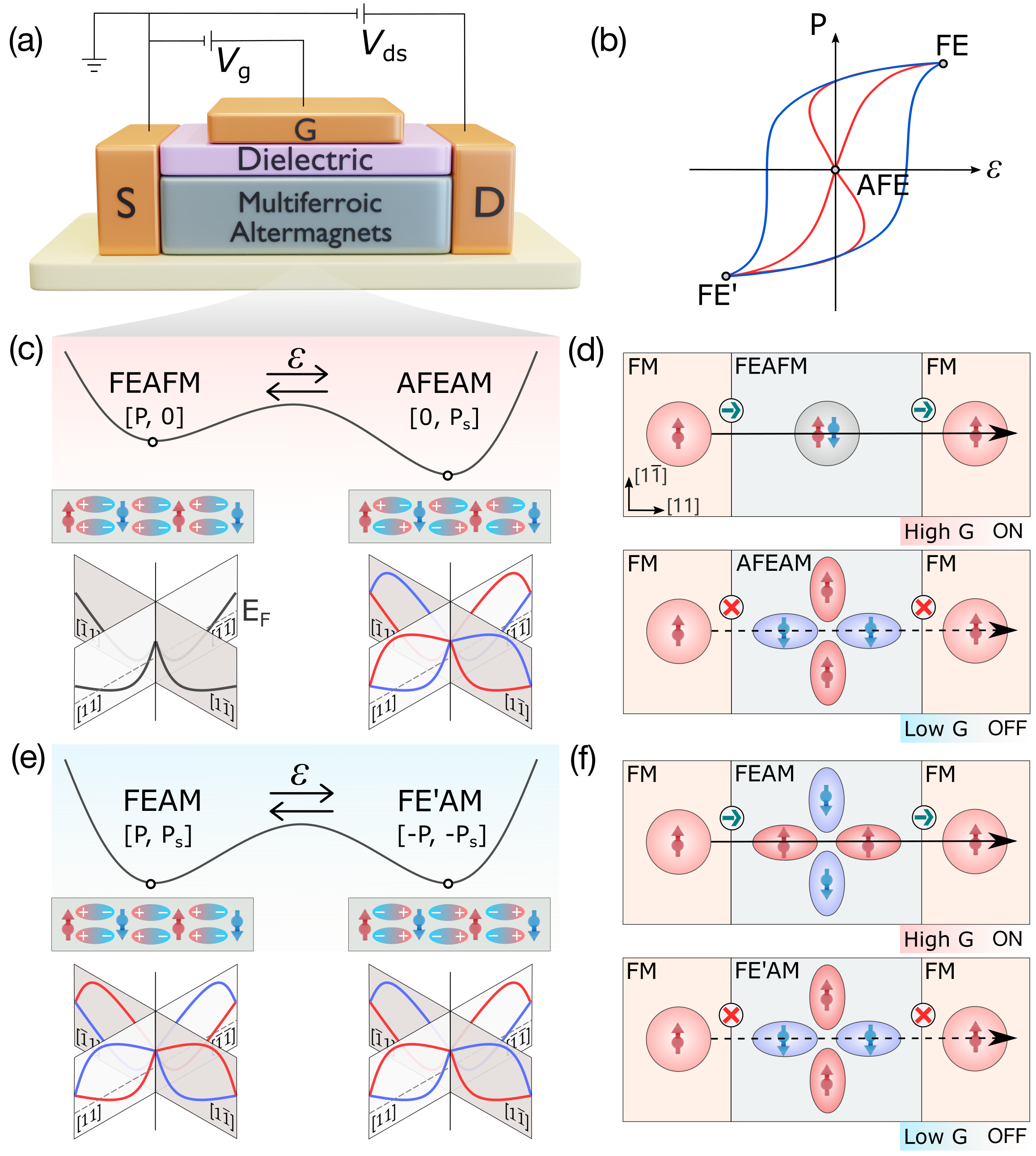}
\caption{ {\bf (a)} AMSFET schematic. {\bf (b)} Hysteretic switching of ferroelectric polarization $\bf P$ in AFE (red) and FE (blue) transitions by electric field $\varepsilon$. {\bf (c)} Schematic configurations of $\bf P$ (ellipses labeled with +/-) and spin polarization $\bf {P_s}$ (red and blue arrows), along with the corresponding illustrative momentum-dependent band structures for the FEAFM and AFEAM states. The gray dashed lines in the bands represent the determined Fermi level, $E_F$. {\bf (d)} Operational principle for FEAFM with high conductance ($G$) and AFEAM with low $G$. The black, red, and blue circles represent the degenerate, spin-up, and spin-down channels, respectively. {\bf (e)} and {\bf (f)} Same as {\bf (c)} and {\bf (d)} but for FEAM-FE$^\prime$AM transition. }
\label{Figure1}
\end{figure}

The proposed AMSFET adopts a conventional three-terminal configuration, as illustrated in Fig.~\ref{Figure1}(a). The source (S) and drain (D) electrodes are modeled as ferromagnetic contacts with fully spin-polarized carriers. A drain–source bias $V_{ds}$ drives current through the multiferroic altermagnet channel, while the gate voltage $V_g$ regulates its ferroelectric phase transition. This, in turn, determines the spin-texture matching between the barrier and the electrodes, thereby modulating spin-dependent transmission and enabling distinct high- and low-conductance states.

The performance of the AMSFET hinges on the properties of the multiferroic altermagnet, which can be broadly categorized into two types of candidate materials: (i) Antiferroelectric altermagnets (AFEAM), include 2D van der Waals compounds such as CuMP$_2$X$_6$ (M = Mo, W; X = S, Se) and the perovskite BiCrO$_3$~\cite{Duan2024PRL}. In these materials, the altermagnetic effect switches the system between a ferroelectric, spin-unpolarized state and an antiferroelectric, spin-polarized state, described by the transition $[\mathbf{P}, 0] \rightleftharpoons [0, \mathbf{P_S}]$. (ii) Ferroelectric altermagnets (FEAM), comprises materials such as Ca$_3$Mn$_2$O$_7$~\cite{Gu2024PRL, Smejkal2024Arxiv}, VXY$_2$ (X = O, S; Y = Cl, Br, I)~\cite{Zhu2025NanoLetters}, and CrPS$_3$~\cite{Wang2025NanoLetters}. In this class, the ferroelectric and spin polarizations reverse simultaneously under switching, following $[\mathbf{P}, \mathbf{P_S}] \rightleftharpoons [-\mathbf{P}, -\mathbf{P_S}]$. These behaviors correspond to the AFE–FE and FE–FE$^\prime$ transitions, respectively, as depicted in the characteristic $P$–$\varepsilon$ hysteresis loops in Fig.~\ref{Figure1}(b).

In the AFEAM–FEAFM transition, the two phases occupy distinct minima on the potential energy surface and can be encoded as binary ``0'' (AFE) and ``1'' (FE), with the AFEAM phase corresponding to the global minimum [Fig.~\ref{Figure1}(c)]. In the FE phase, the system retains conventional AFM order, leading to spin-degenerate bands throughout the Brillouin zone. In contrast, the AFE phase exhibits AM order with momentum-dependent spin splitting. Along the [11] direction, only spin-down carriers remain at the Fermi level, yielding 100\% spin polarization [Fig.~\ref{Figure1}(c)]. This polarization switching leads to significant conductance contrast in the AMSFET. As shown in Fig.~\ref{Figure1}(d), when the barrier is in the FE state, its spin texture matches that of the electrodes, enabling efficient spin-up transport and resulting in a high-conductance ON state. When switched to the AFE state, however, the spin texture becomes mismatched, blocking spin-up electrons due to a lack of available states at the same momentum. This mismatch suppresses current flow, producing a low-conductance OFF state. 

The FEAM–FE$^\prime$AM transition features similarly but is governed by a double-well potential with two equivalent ferroelectric states, labeled ``$+1$'' and ``$-1$'' [Fig.~\ref{Figure1}(e)]. Reversing the ferroelectric polarization simultaneously flips the spin polarization of the altermagnets, such that the FE and FE$^\prime$ states support 100\% spin-up and spin-down conduction along [11] direction, respectively. As depicted in Fig.~\ref{Figure1}(f), this reversal toggles the spin alignment between the barrier and the electrodes, amplifying the conductance contrast between the ON and OFF states. This spin-selective transport mechanism is the basis of reliable binary switching in the AMSFET.

\begin{figure*}[t]
	\includegraphics
	[width=0.94\textwidth]{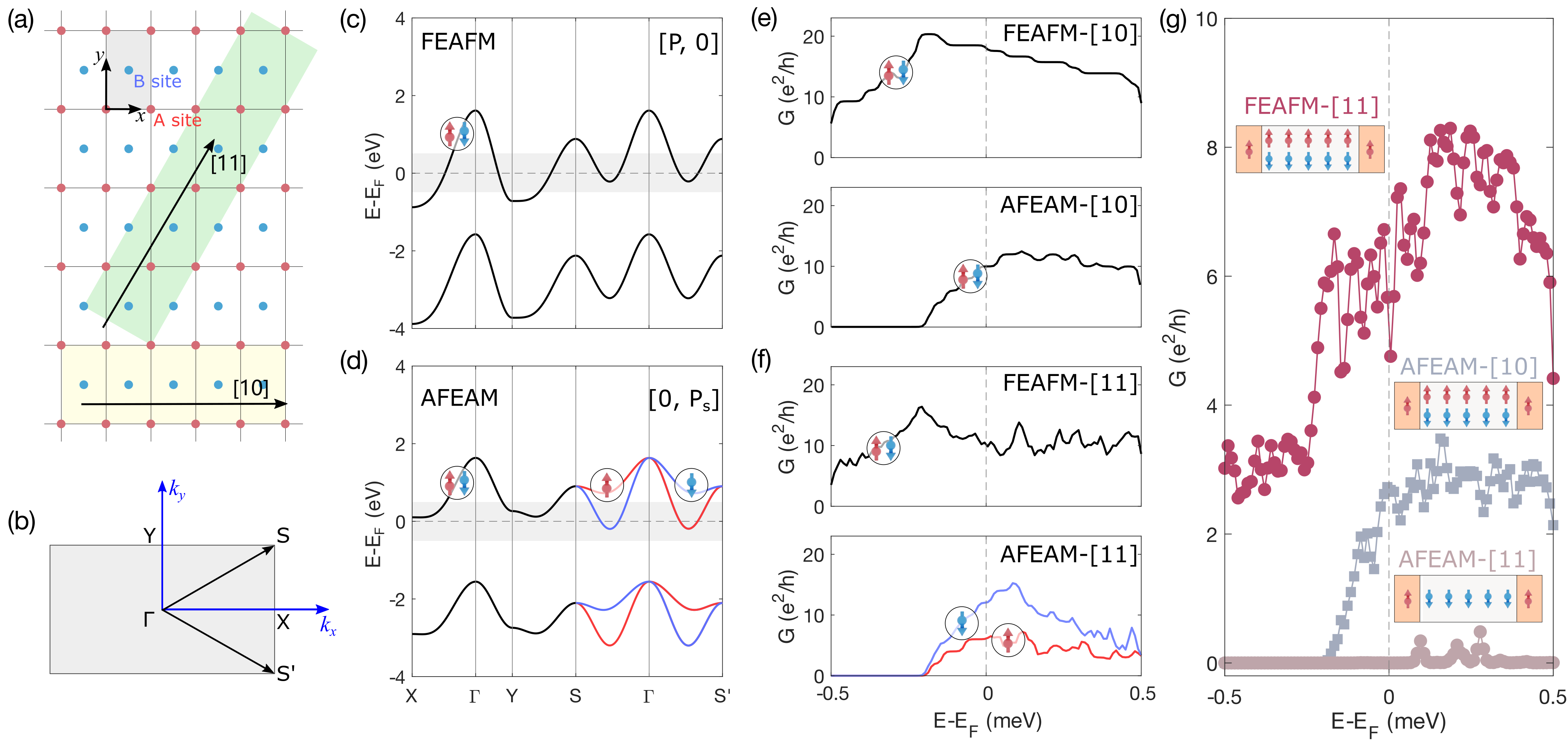}
	\caption{ {\bf(a)} Schematic illustration of 2D rectangular magnetic sublattices $A$ (red) and $B$ (blue) with nested AFM lattice. The gray shadow represents the unit cell, the yellow and green shadows represent the [11] and [10] orientations of the calculated nanoribbon model. {\bf(b)} First Brillouin zone with high symmetry points. {\bf(c)} Spin-resolved band structures for FEAFM and {\bf(d)} AFEAM state calculated from TB model. The gray section represents the energy region for calculating conductance ($G$). {\bf(e)} Spin-dependent $G$ as a function of energy calculated along [10] and {\bf(f)} [11] direction for FEAFM and AFEAM states, respectively. {\bf(g)} Total $G$ of the AMSFET device with two leads. The inset illustrates the corresponding spin configuration and lattice orientation.}
	\label{Figure2}
\end{figure*}

To illustrate the design principle, we construct a 2D tight-binding (TB) model to describe the electronic states of multiferroic altermagnets and calculate their conductance. Figure~\ref{Figure2}(a) illustrates our setup on a general 2D rectangular lattice. The model Hamiltonian expressed as

\begin{equation}
	\begin{aligned}
		H &= \sum_{i,j} \left( 
		f_i^{\boldsymbol{\eta}_j} \, c_i^\dagger c_{i+\boldsymbol{\eta}_j} 
		+ g_i^{\boldsymbol{\kappa}_j} \, c_i^\dagger c_{i+\boldsymbol{\kappa}_j} 
		+ h_i^{\boldsymbol{\delta}_j} \, c_i^\dagger c_{i+\boldsymbol{\delta}_j} 
		+ l_i^{\boldsymbol{\gamma}_j} \, c_i^\dagger c_{i+\boldsymbol{\gamma}_j} 
		\right.\\ 
		&\left. \quad + \text{H.C.} \right) + M_{A,B} \sum_{i \in A,B} c_i^\dagger \sigma_z c_i.
		\label{equ1}
	\end{aligned}
\end{equation}

\noindent Here, $c_i^{\dagger}$($c_i$) are electron creation (annihilation) operator at site $i$ and $\sigma$ is the Pauli matrix. The parameters $f_i^{\boldsymbol{\eta}_j}$, $g_i^{\boldsymbol{\kappa}_j}$, $h_i^{\boldsymbol{\delta}_j}$, and $l_i^{\boldsymbol{\gamma}_j}$ describe the electron hopping between site $i$ and its first- (NN), second- (2NN), third- (3NN), and fourth-nearest neighbors (4NN), connected by the vectors $\boldsymbol{\eta}_j$, $\boldsymbol{\kappa}_j$, $\boldsymbol{\delta}_j$, and $\boldsymbol{\gamma}_j$, respectively. The AFM exchange field is characterized by $M_{A,B}$ on sublattices $A$ and $B$, with $M_{A} = -M_{B}$. Detailed parameters are provided in the Supplemental Material (SM)~\cite{SM}.

The emergence of altermagnetism is determined by the equivalence of hopping parameters between two spin-opposed sublattices~\cite{Duan2024PRL, Zhu2025NanoLetters}. For the FEAFM–AFEAM transition, the equivalence of the 3NN hopping terms in Eq.~\eqref{equ1} is crucial~\cite{Duan2024PRL}. In this case, the FE phase exhibits a conventional AFM state with spin-degenerate bands across the Brillouin zone [Fig.~\ref{Figure2}(c)], whereas the AFE phase manifests as an AM state characterized by momentum-dependent spin splitting [Fig.~\ref{Figure2}(d)].

For a device with two attached spin-polarized ferromagnetic electrodes, we calculated the spin-resolved conductance using the Landauer-B{\"u}ttiker formalism \cite{Datta1995:Book} and Green's-function method~\cite{Sancho1984:JPFMP,Jiang2009:PRB,Qiao2016:PRL,Cheng2018:PRB}. We consider the device with different ferroelectric states along multiple crystal directions, and the transport setup and detailed parameters are provided in SM~\cite{SM}. Along the [10] crystal direction, both AFEAM and FEAFM exhibit spin-degenerate states, the spin-up and spin-down channels carry charge carriers equally, resulting in comparable conductance, as shown in Fig.~\ref{Figure2}(e). In contrast, along the [11] crystal direction for AFEAM, the momentum-dependent spin splitting characteristic of the altermagnetic phase induces an asymmetry between spin-up and spin-down channels, leading to a pronounced conductance difference [Fig.~\ref{Figure2}(f)].

To realize the SFET through altermagnetic properties, we first consider FEAFM and AFEAM along [11] crystal direction. For the specific spin-polarized ferroelectric electrodes, i.e., spin-up, charge carriers are efficiently transmitted for the spin-degenerate conduction channels in FEAFM, producing a high-conductance [Fig.~\ref{Figure2}(g)], denoting ``ON'' state. On the contrary, the transport channels in AFEAM are spin-down [Fig.~\ref{Figure2}(d)], resulting in backscattering for spin-up charge carriers, leading to a low-conductance [Fig.~\ref{Figure2}(g)], which corresponds to ``OFF'' state. The switching between FEAFM and AFEAM can be realized electrically~\cite{Duan2024PRL}, thus a gate controlled AMSFET is realized. Notably, the conductance in the AFEAM state approaches zero, corresponding to an effectively infinite ON/OFF ratio. Furthermore, the difference between AFEAM conductance along [10] and [11] underscores the importance of lattice orientation, providing an additional degree of freedom for device optimization [Fig.~\ref{Figure2}(g)].

\begin{figure}[t]
	\includegraphics
	[width=0.48\textwidth]{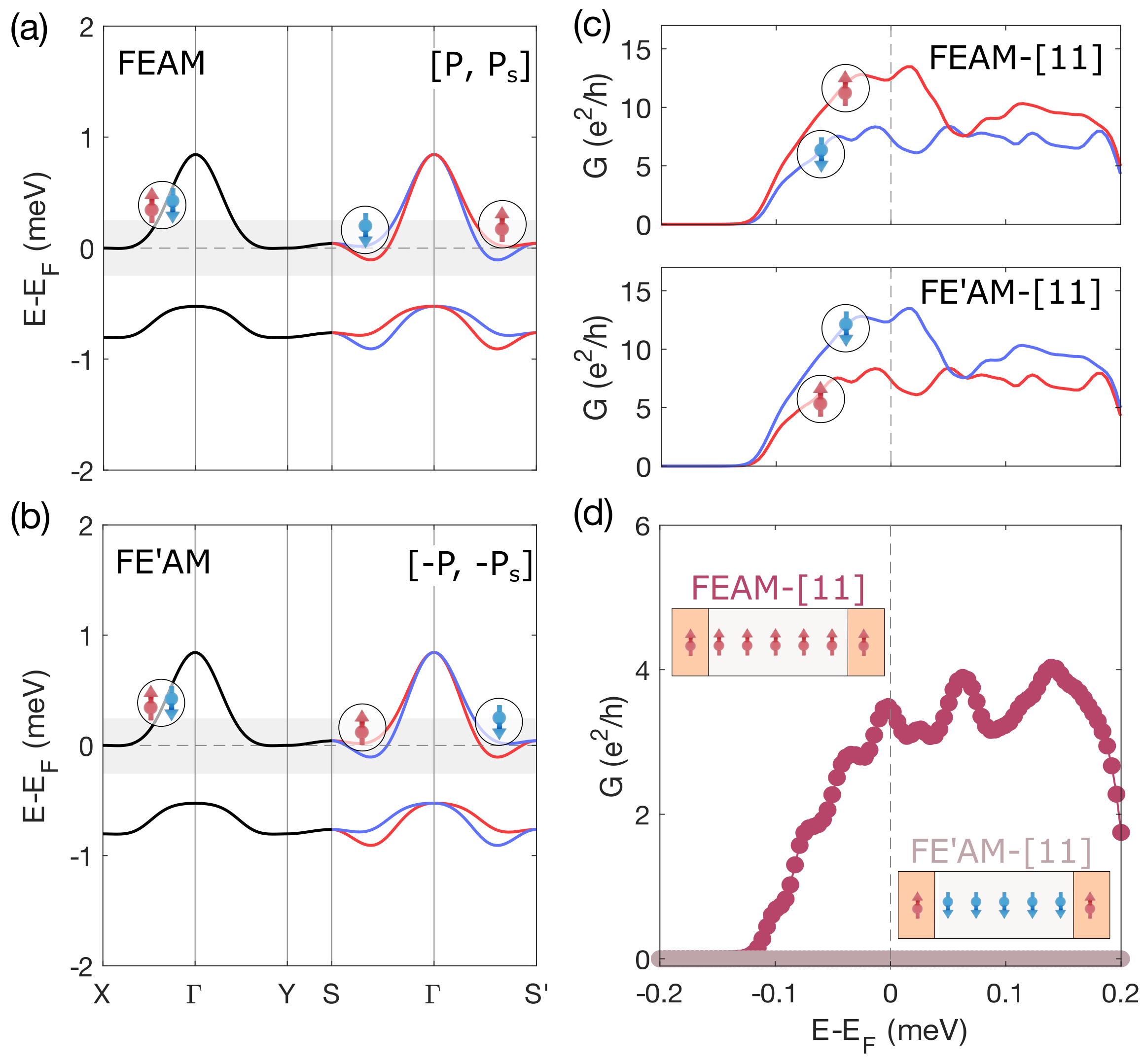}
	\caption{ {\bf(a)} Spin-resolved band structures for FEAM and {\bf(b)} FE$^\prime$AM state calculated from TB model. {\bf(c)} Spin-dependent conductance as a function of energy calculated along [11] direction for FEAM and FE$^\prime$AM states, respectively. {\bf(d)} Total $G$ of the AMSFET device with leads. The inset illustrates the corresponding spin configuration and lattice orientation.}
	\label{Figure3}
\end{figure}

In addition, we propose another route to realizing AMSFET operation by exploiting the spin-polarization reversal along the [11] direction in the FEAM--FE$^\prime$AM transition. The spin-resolved band structures of FEAM [Fig.~\ref{Figure3}(a)] and FE$^\prime$AM [Fig.~\ref{Figure3}(b)] show that reversing the ferroelectric polarization results in a simultaneous flip of the spin polarization, as also verified in Ref.~\cite{Zhu2025NanoLetters}. The flip of the spin polarization of the transport channels changes the spin polarization of conductance, as shown in Fig.~\ref{Figure3}(c). For the spin-up charge carriers, FEAM opens the spin-up transport channels while FE$^\prime$AM opens the spin-down channels, thus resulting in high-conductance (``ON'' state) and low-conductance (``OFF'' state) [Fig.~\ref{Figure3}(d)], respectively. This also confirms the realization of an AMSFET within AM.

\begin{figure}[t]
	\includegraphics
	[width=0.48\textwidth]{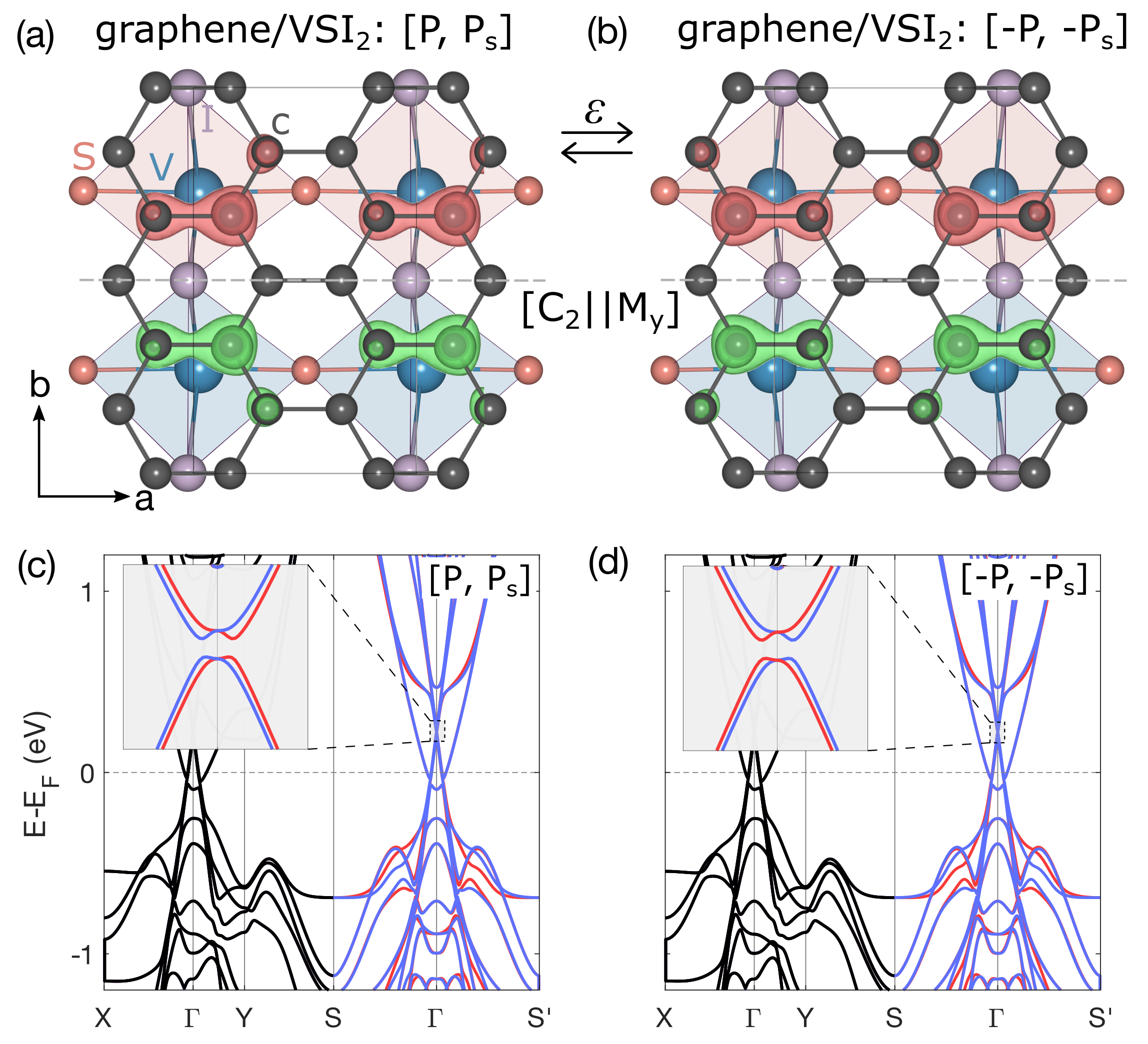}
	\caption{ {\bf(a)} and {\bf(b)} Crystal structure of graphene/VSI$_2$ heterostructure in opposite ferroelectric (FE) phases. Red and blue polyhedra represent the two opposite spin sublattices, while the spin densities induced in graphene are highlighted in red and green, the isosurface value is $3 \times 10^{-5}$~e/bohr$^3$. {\bf(c)} and {\bf(d)} Corresponding calculated band structures, where black, red, and blue lines denote spin-degenerate, spin-up, and spin-down bands, respectively. Inset: band structure near the Dirac point of the graphene component. }
	\label{Figure4}
\end{figure}


A key challenge in integrating multiferroics into a spin field-effect transistor arises from the intrinsic incompatibility between metallic conductivity and ferroelectricity. Specifically, mobile carriers driven by $V_{ds}$ can strongly screen the external electric field, hindering polarization reversal, and highly leaky samples typically fail to produce a reliable $P$–$\varepsilon$ hysteresis loop~\cite{Lines2001book, Kim2016APL}. In the proposed AMSFET architecture, this limitation is overcome via the altermagnetic proximity effect (AMPE) [Fig.~\ref{Figure1}(a)]~\cite{Zhu2025Arxiv, Zutic2019MT}. To construct the transport channel, a nonmagnetic layer is deposited atop the multiferroic altermagnetic layer. Through the AMPE, this nonmagnetic layer inherits the momentum-dependent spin texture of the underlying substrate, effectively forming a proximitized altermagnet (PAM) layer. As a result, the ferroelectric polarization is confined to the insulating multiferroic altermagnetic layer, while metallic conduction is localized in the PAM layer, achieving spatial separation yet functional coupling.

As a preliminary step toward verifying the AMPE in multiferroics, we performed first-principles calculations on a graphene/VSI$_2$ heterostructure, combining the prototypical 2D material graphene with the FEAM VSI$_2$~\cite{Zhu2025NanoLetters}. VSI$_2$ crystallizes in an orthorhombic lattice in which two inequivalent V sublattices are related by a mirror symmetry $M_y$, arising from lattice distortions due to pseudo–Jahn–Teller instabilities and Peierls-like dimerization~\cite{Zhu2025NanoLetters, Zhang2021PRB}. This symmetry underlies the hallmark features of altermagnetism, as confirmed by staggered spin densities in real space and momentum-dependent spin splitting (see Supplemental Material~\cite{SM}).

In the graphene/VSI$_2$ heterostructure, the layer-resolved band structure indicates that each constituent largely retains its intrinsic electronic character under van der Waals coupling (Supplemental Material~\cite{SM}). Remarkably, graphene acquires pronounced momentum-dependent spin splitting [Fig.~\ref{Figure4}(c)], in stark contrast to its pristine spin-degenerate state, establishing it as a proximity-induced altermagnetic (PAM) state via the AMPE. The induced spin texture follows the characteristic altermagnetic pattern: spin degeneracy is preserved along X--\(\Gamma\)--Y--S, while opposite spin polarizations appear along S--\(\Gamma\)--S$^\prime$, consistent with the underlying symmetry of VSI$_2$. Notably, the band gap opening in graphene originates from the staggered potential imposed by the substrate, and the real-space spin density in graphene mirrors the symmetry of the VSI$_2$ layer, confirming the inheritance of altermagnetic character, as shown in Fig.~\ref{Figure4}(a).

Crucially, when the ferroelectric state of VSI$_2$ is switched, with V atoms displacing along the $a$-axis either to the left (FE) or to the right (FE$^\prime$) within the [VS$_2$I$_4$] octahedra, the spin polarization of graphene's bands reverses accordingly [Figs.~\ref{Figure4}(c) and~\ref{Figure4}(d)]. This demonstrates that in an AMSFET based on graphene/VSI$_2$ heterostructure, an external electric field that reverses the ferroelectric state of VSI$_2$ induces a spin flip in the PAM-graphene layer along defined orientations. Such a mechanism enables fully electrical control of spin-dependent conductance and highlights graphene/VSI$_2$ as a strong candidate for AMSFET applications.

Our results position the AMSFET as a fundamentally new platform for spintronic devices. It breaks the conventional paradigm of SFET architectures while simultaneously resolving the intrinsic conflict between ferroelectric insulation and metallic transport in multiferroics. Antiferromagnet-based systems, in particular, exhibit terahertz-scale dynamics and produce no stray magnetic fields, thereby enhancing the energy efficiency and speed of magnetic switching. The underlying control mechanism originates from non-relativistic effects and does not require reversal of the N$\acute{\text{e}}$el vector, enabling straightforward electric control of spin polarization. Furthermore, integration with conventional CMOS platforms may enable hybrid spintronic–electronic circuits that exploit nonvolatile spin control for memory and logic applications. Moreover, the ability to electrically modulate momentum-dependent spin textures opens a pathway to more complex functionalities, including spin filtering, reconfigurable logic, and topological spin transport, potentially enabling device operation beyond binary switching toward multi-level information processing. These features establish the AMSFET as a promising platform for next-generation electrically controlled spintronic devices.


\medskip
\indent{$Acknowledgements$}---This work is supported by the National Natural Science Foundation of China (12474155, 12447163, 12504108, and 11904250), the Zhejiang Provincial Natural Science Foundation of China (LR25A040001), the China Postdoctoral Science Foundation (2025M773440), and the U.S. DOE, Office of Science BES, Award No. DE-SC0004890 (I.\v{Z}.). The computational resources for this research were provided by the High Performance Computing Platform at the Eastern Institute of Technology, Ningbo.

Z.Z. and X.C. contributed equally to this work.

\bibliography{reference}

\end{document}